\documentclass[aps, prl, amsmath, showpacs, preprintnumbers, twocolumn, amssymb, amsmath, superscriptaddress]{revtex4}

\usepackage{graphicx}
\usepackage{mathptmx, textcomp}
\usepackage[latin1]{inputenc}
\usepackage{tabularx}
\usepackage{units}

\begin{document}
\title{Narrow-line magneto-optical trap for erbium}
\author{A. Frisch}
\affiliation{Institut f\"ur Experimentalphysik and Zentrum f\"ur Quantenphysik, Universit\"at Innsbruck, Technikerstra{\ss}e 25, 6020 Innsbruck, Austria}
\author{K. Aikawa}
\affiliation{Institut f\"ur Experimentalphysik and Zentrum f\"ur Quantenphysik, Universit\"at Innsbruck, Technikerstra{\ss}e 25, 6020 Innsbruck, Austria}
\author{M. Mark}
\affiliation{Institut f\"ur Experimentalphysik and Zentrum f\"ur Quantenphysik, Universit\"at Innsbruck, Technikerstra{\ss}e 25, 6020 Innsbruck, Austria}
\author{A. Rietzler}
\affiliation{Institut f\"ur Experimentalphysik and Zentrum f\"ur Quantenphysik, Universit\"at Innsbruck, Technikerstra{\ss}e 25, 6020 Innsbruck, Austria}
\author{J. Schindler}
\affiliation{Institut f\"ur Experimentalphysik and Zentrum f\"ur Quantenphysik, Universit\"at Innsbruck, Technikerstra{\ss}e 25, 6020 Innsbruck, Austria}
\author{E. Zupani\v{c}}
\affiliation{Institut f\"ur Experimentalphysik and Zentrum f\"ur Quantenphysik, Universit\"at Innsbruck, Technikerstra{\ss}e 25, 6020 Innsbruck, Austria}
\affiliation{Jo\v{z}ef Stefan Institute, Jamova 39, SI-1000 Ljubljana, Slovenia}
\author{R. Grimm}
\affiliation{Institut f\"ur Experimentalphysik and Zentrum f\"ur Quantenphysik, Universit\"at Innsbruck, Technikerstra{\ss}e 25, 6020 Innsbruck, Austria}
\affiliation{Institut f\"ur Quantenoptik und Quanteninformation,
 \"Osterreichische Akademie der Wissenschaften, 6020 Innsbruck, Austria}
\author{F. Ferlaino}
\affiliation{Institut f\"ur Experimentalphysik and Zentrum f\"ur Quantenphysik, Universit\"at Innsbruck, Technikerstra{\ss}e 25, 6020 Innsbruck, Austria}

\date{\today}


\begin{abstract}
We report on the experimental realization of a robust and efficient magneto-optical trap for erbium atoms, based on a narrow cooling transition at 583~nm.  We observe up to $N=2 \times 10^{8}$ atoms at a temperature of about $T=\unit[15]{\mu K}$. This simple scheme provides better starting conditions for direct loading of dipole traps as compared to approaches based on the strong cooling transition alone, or on a combination of a strong and a narrow kHz transition. Our results on Er point to a general, simple and efficient approach to laser cool samples of other lanthanide atoms (Ho, Dy, and Tm) for the production of quantum-degenerate samples.
\end{abstract}

\maketitle

\begin{figure}[t]
\includegraphics[width=0.9\columnwidth] {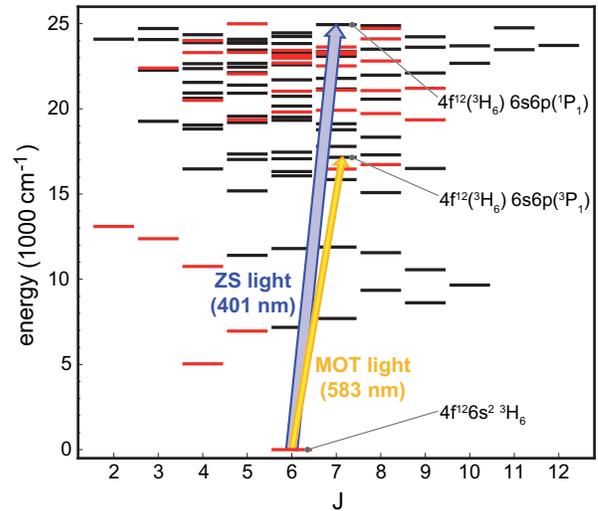}
\caption{(color online). Energy levels of atomic erbium up to $\unit[25 000]{cm^{-1}}$ for different total electronic angular momentum quantum numbers $J$ \cite{Erbium}. States with odd (even) parity are indicated by black (red) horizontal lines. The two relevant laser cooling transitions at 401 and 583~nm are indicated by arrows \cite{Ban2005lct}.}
\label{fig:levelscheme}
\end{figure}

Laser cooling of non-alkali atoms has become a very active and challenging field of research. The great appeal of  unconventional atomic systems for experiments on ultracold atomic quantum gases stems from the possibility of engineering  complex interactions and of accessing rich atomic energy spectra. Both features are at the foundation of a number of novel fascinating phenomena. For instance the energy spectra of  two-valence-electron species, as alkaline earth and alkaline-earth-like atoms, feature narrow and ultra-narrow optical transitions, which are key ingredients for ultra-precise atomic clocks \cite{Ido2003rfs}, efficient quantum computation schemes \cite{Daley2008qcw}, and novel  laser cooling approaches as beautifully demonstrated in experiments with Sr, Yb, and Ca  \cite{Katori1999mot,Kuwamoto1999mot, Curtis2001qnl}.

As a next step in complexity, multi-valence-electron atoms with non-$S$ electronic ground state such as lanthanides are currently attracting an increasing experimental and theoretical interest. Among many, one of the special features of lanthanides is the exceptionally large magnetic dipole moment of atoms in the electronic ground state (e.\,g.\,$\unit[7]{\mu_B}$ for Er and \,$\unit[10]{\mu_B}$ for both Dy and Tb), which provides a unique chance to study strongly dipolar phenomena with atoms. Highly magnetic atoms  interact with each other not only via the usual contact interaction but also via an anisotropic and long-range interaction, known as the dipole-dipole interaction \cite{Lahaye2009tpo}.  Chromium was the first atomic species used for experiments on atomic dipolar quantum gases \cite{Griesmaier2005bec,Beaufils2008aop}, and  the even more magnetic lanthanides are nowadays in the limelight thanks to laser cooling experiments on Er and Tm \cite{Mcclelland2006lcw,Sukachev2010mto} and to the recent realization of quantum-degenerate Dy gases \cite{Lu2011sdb,Lu2012qdd}.

Similarly to Yb and the alkaline earth atoms, the atomic energy spectra of magnetic lanthanides include broad, narrow, and ultra-narrow optical transitions. This collection of lines is reflected in a wide choice of possible schemes for laser cooling experiments. However, all experiments on Zeeman slowing and cooling in a magneto-optical trap (MOT) with magnetic lanthanides so far relied  on an approach, essentially based on the strongest cycling transition  \cite{Mcclelland2006lcw,Sukachev2010mto,Youn2010dmo}. This broad transition typically lies in the \emph{blue} between 400 and 430~nm and has a linewidth on the order of few tens of MHz. As a consequence, the Doppler temperature is close to a mK. Such a high temperature makes direct loading from a MOT into a dipole trap difficult and inefficient, even when sub-Doppler mechanisms take place \cite{Berglund2008nlm, Sukachev2011sdl}. To further decrease the temperature of atoms prior to the dipole trap loading, an additional MOT stage based on an ultra-narrow kHz-linewidth  transition was applied in Refs.\,\cite{Youn2010dmo,Lu2011sdb}, making the whole experimental procedure more involved.

Taking advantage of the rich atomic spectrum of lanthanides, we identify a different transition to be the most suitable one for MOT operation towards production of a quantum-degenerate gas. This transition, which for Er lies at 583~nm and has a linewidth of  190~kHz, represents an intermediate case between the broad (blue) and the ultra-narrow (red) transitions  available in the lanthanide spectra and has very similar properties to the Yb intercombination line \cite{Takasu2003hdt,Takasu2003ssb}. Based on this narrow yellow  line, we demonstrate a MOT of $2\times 10^8$ Er atoms at a temperature as low as $\unit[15]{\mu K}$. Our  approach, inspired by Yb experiments \cite{Kuwamoto1999mot}, provides better starting conditions for direct loading into a dipole trap than the conditions achieved in other experiments with magnetic lanthanides \cite{Lu2011sdb}.

Erbium is a heavy rare-earth element of the lanthanide series. It has six stable isotopes, among which three bosonic isotopes $^{166}$Er ($34\%$), $^{168}$Er ($27\%$), and $^{170}$Er ($15\%$) and the fermionic isotope $^{167}$Er ($23\%$) have high natural abundance.
The Er electronic configuration is characterized by a xenon-like core, an inner open $4f$ shell, and an outer closed $6s$ shell, $[\rm{Xe}]4f^{12}6s^{2}$. The electron vacancies in the inner $f$ shell are a common feature of all the lanthanides (with the exception of Yb) and are at the origin of the strong magnetism as well as various interesting collisional effects \cite{Krems2004eia,Connolly2010lsr,Kotochigova2011ait}.

The atomic level spectrum of Er is shown in  Fig.\,\ref{fig:levelscheme}. In the ground state, Er has a highly anisotropic electronic density distribution with a large orbital angular momentum $L=5$ ($H$-state) and a total electronic angular momentum $J=6$. The bosonic isotopes have zero nuclear spin ($I=0$) and consequently do not exhibit a hyperfine structure. On the contrary, the fermionic isotope $^{167}$Er has  a nuclear spin $I=7/2$, leading to eight hyperfine levels, from $F=5/2$ to $F=19/2$, in the electronic ground state.

In Ref.\,\cite{Ban2005lct}, five different $J\rightarrow J+1$ laser cooling transitions were identified with linewidths ranging from tens of MHz to a few Hz. Here we focus on the blue and the yellow transition at 401~nm and at 583~nm, respectively; see arrows in Fig.\,\ref{fig:levelscheme}. The corresponding excited levels are the singlet $^1P_1$ and triplet $^3P_1$ states coming from the transition of an  $s$-electron into a $p$-shell. The strong blue transition at $\unit[401]{nm}$ has a linewidth of $\Gamma_{401}/2\pi= \unit[27.5]{MHz}$ \cite{Note1}, corresponding to a Doppler temperature of   $\hbar \Gamma_{401}/(2 k_{\rm B})=\unit[660]{\mu K}$. We use the blue light for both  Zeeman slowing and transversal cooling of the Er atomic beam. The MOT operates on the narrow yellow transition at $\unit[583]{nm}$, which has a natural linewidth of  $\Gamma_{583}/2\pi= \unit[190]{kHz}$  and a corresponding Doppler temperature of $\unit[4.6]{\mu K}$. The line may exhibit weak leaks from the  excited level into two intermediate levels with very low calculated leakage rates (0.017\,s$^{-1}$ and 0.0049\,s$^{-1}$) \cite{Ban2005lct}. Note that we find these losses irrelevant for all practical purposes. For $^{167}\mathrm{Er}$ the hyperfine structure in the 401- and 583-nm excited states gives rise to eight levels ranging from $F'=7/2$ to $F'=21/2$. For the 583-nm line, the hyperfine constants are known \cite{Childs1983hso}. Contrary, the hyperfine constants for the 401-nm line are currently unknown, making the operation of the fermionic MOT more challenging.

We generate the blue light from two independent blue diode lasers injection locked to a master laser. The master laser light is produced by frequency doubling methods based on a tapered amplified diode laser at $\unit[802]{nm}$. The blue light from the master laser is locked to a hollow-cathode discharge lamp \cite{Kuwamoto1999mot} via modulation transfer spectroscopy. With this setup we  spectroscopically resolve the lines of the four most abundant bosonic isotopes as well as the hyperfine structure of the fermionic isotope $^{167}\mathrm{Er}$. Since the hyperfine constants of the excited level at $\unit[401]{nm}$ are unknown we could not assign the absorption features to specific hyperfine transitions. We derive the yellow light from a dye laser operating with Rhodamin 6G. By using an intracavity EOM and an external reference cavity, we stabilize the laser to a linewidth of about $\unit[50]{kHz}$. By additionally locking the laser to an ultra-low-expansion cavity, we achieve a long-term stability better than $\unit[30]{kHz}$ within a day \cite{Note3}.

Our experimental procedure is as follows. We load the Zeeman slowed atomic beam from an effusive, high-temperature oven directly into the narrow-line MOT. Our commercial oven typically operates at a temperature of $\unit[1300]{^\circ C}$, which is about $\unit[200]{^\circ C}$ below the Er melting point. Two $3$-mm apertures, separated by $\unit[50]{mm}$, provide a geometrical collimation of the atomic beam. In addition, the atomic beam is further collimated and transversally cooled by a 2D optical molasses, working on the broad 401-nm transition with a total power of about $\unit[100]{mW}$. The beams are elliptically shaped to increase the interaction time between the atoms and the light. Thanks to the transversal cooling stage, we increase the loading flux by almost an order of magnitude.

The atomic beam then enters the Zeeman slower (ZS).
Because of the limited capture velocity (a few $\unit[]{m/s}$) imposed by the narrow cooling transition used for the MOT, it is crucial to design a ZS that provides enough atomic flux at low velocities \cite{Kuwamoto1999mot}. We build a $\unit[360]{mm}$ long spin-flip Zeeman slower, which can slow the atoms from $\unit[500]{m/s}$ to about $\unit[5]{m/s}$.
The ZS light is focused at the oven position and has a total power of about $\unit[60]{mW}$. At the MOT position, we estimate a beam diameter of about $\unit[11]{mm}$, corresponding to an intensity of about one ${I_{s, 401}}$, where $\unit[]{I_{s, 401}}=\unit[56]{mW/cm^2}$ is the saturation intensity. Our ZS operates with light detuned by about $\unit[-20]{\Gamma_{401}}$ ($\unit[-540]{MHz}$) from the unshifted resonance.

The narrow-line MOT is operated in a standard six beam configuration with retro-reflected beams. To increase the capture velocity of the MOT, we use  large MOT beam diameters of about $\unit[30]{mm}$. Typical MOT loading parameters include a magnetic field gradient $B'$ (along the symmetry axis) of $\unit[4]{G/cm}$, a laser intensity of $\unit[12]{I_{s, 583}}$ per beam with ${\rm I_{s, 583}}=\unit[0.13]{mW/cm^2}$, and a detuning $\delta_{583}$ from the atomic transition of $\unit[-50]{\Gamma_{583}}$ ($\unit[-9.5]{MHz}$). To measure the number of atoms in the MOT after loading, we optically compress the MOT by reducing $\delta_{583}$ to $\unit[-0.5]{\Gamma_{583}}$ and we apply standard absorption imaging on the blue transition.

\begin{figure}
\includegraphics[width=1\columnwidth] {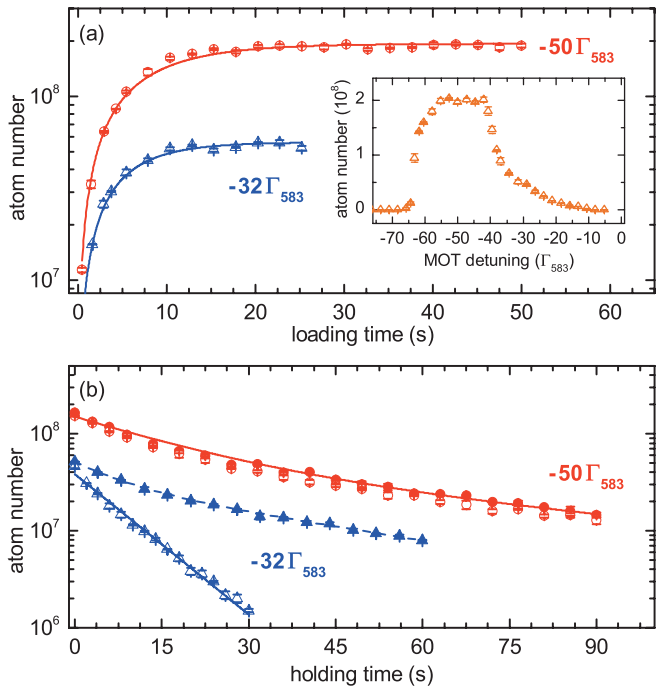}
\caption{(color online). Loading (a) and decay (b) of the narrow-line MOT with $^{166}$Er atoms for $\delta_{583}=-50\Gamma_{583}$ (circles) and $\delta_{583}=-32\Gamma_{583}$ (triangles). (a) Atom numbers are plotted as a function of the MOT loading time. The solid lines are fits to the data using $N(t)=N_{\rm ss}(1-e^{-\gamma t})$ with $N_{\rm ss}=R/\gamma$. From the fits we obtain $\gamma=\unit[0.137(2)]{s^{-1}}$, $R=\unit[2.6(4)\times 10^7]{s^{-1}}$, and $\gamma=\unit[0.200(4)]{s^{-1}}$, $R=\unit[1.1(1)\times 10^7]{s^{-1}}$ for $-50\Gamma_{583}$ and $-32\Gamma_{583}$, respectively. The inset shows the atom number after $\unit[10]{s}$ loading time as a function of the MOT detuning. (b) Atom numbers are plotted as a function of the holding time in the MOT in presence (empty symbols) and absence of the ZS light (filled symbols). The solid lines are fits to the data using a double exponential function; see text.}
\label{fig:MOTloading}
\end{figure}

A special feature of our narrow-line MOT is the large detuning of the MOT light (typically $\unit[-50]{\Gamma_{583}}$) needed for optimal loading.
At this detuning, we observe a very strong effect of gravity on the position and shape of the atom cloud \cite{Katori1999mot}. The atoms are located well below the center of the magnetic quadrupole field, and the cloud takes the form of a large semi-shell. To elucidate the  reason for the large detuning, we monitor the loading dynamics and the lifetime of the MOT; see Fig.\,\ref{fig:MOTloading}. Our measurements focus on the $^{166}$Er isotope, but we have observed the same qualitative behavior also for the other isotopes.

Figure \ref{fig:MOTloading}(a) shows the atom number in the MOT as a function of the loading time for different  values of $\delta_{583}$. We fit our data by using a standard loading rate equation \cite{Weiner1999eat}, which includes a capture rate $R$ and a decay rate $\gamma$; the latter accounts for both collisions between trapped atoms and collisions with the background gas.
For a detuning of $\unit[-50]{\Gamma_{583}}$ the atom number approaches its steady state in about $\unit[10]{s}$ with $N_{\rm ss}\approx 2\times 10^8$.
For a lower value of  the detuning to $\unit[-32]{\Gamma_{583}}$ we observe a substantial decrease of the atom number to $N_{\rm ss}\approx 5\times 10^7$. This behavior is clearly shown in the inset, where we monitor the number of atoms $N_{\rm ss}$ for a fixed loading time of $\unit[10]{s}$ as a function of the MOT light detuning. For detunings exceeding $\delta_{583}=\unit[-30]{\Gamma_{583}}$ we observe a rapid increase in the atom number. When further increasing the detuning, $N_{\rm ss}$ first stays   constant and then rapidly decreases.

We believe that the large detuning of the MOT light serves us to minimize the detrimental effects of off-resonant pumping processes driven by the ZS light. According to the Zeeman shift in the quadrupole field of the MOT,
we observe a spatial displacement of the atomic cloud with $\delta_{583}$ of about $\unit[1.4]{mm/MHz}$ \cite{Note2}. For large detunings, this shift becomes so large that the atoms can no longer be kept in the MOT. For intermediate detunings in the range from $-40$ to $\unit[-55]{\Gamma_{583}}$, the cloud displacement is  advantageous for MOT operation since the atoms become spatially separated from the region of interaction with the ZS light. For small detunings, the ZS light leads to substantial losses of atoms.

The effect of the ZS light also shows up in lifetime measurements, where  we monitor the number of atoms in the MOT as a function of the holding time with the ZS light being present or absent; see Fig.\,\ref{fig:MOTloading}(b). For these measurements we switch off the ZS magnetic field and the atomic beam after $\unit[10]{s}$ of MOT loading. For large detuning ($\delta_{583}=\unit[-50]{\Gamma_{583}}$) the evolution of the atom number is not affected by the ZS light. In both cases, i.\,e.\,with the ZS light on and off, we observe faster losses in the earlier stage of the decay, which we attribute to inelastic two-body collisional processes, and a slower decay at a later time, which is finally limited by background collisions. For simplicity we use a double-exponential fit function to estimate the time constants \cite{Note4}. We extract time constants of $\unit[19(3)]{s}$ and $\unit[80(20)]{s}$ for the fast and slow dynamics, respectively. At lower MOT detuning ($\unit[-32]{\Gamma_{583}}$), the ZS light strongly affects the decay. When the blue light is turned off, our observations are qualitatively similar to the ones at $\unit[-50]{\Gamma_{583}}$ while with the blue light turned on the atomic loss dramatically increases. In this case the decay curve is well described by a single exponential function with a time constant $\tau=\unit[9.0(1)]{s}$. This decay time is consistent with a simple estimate of pumping losses. By considering the absorption rate $\Gamma_a$ of the 401-nm light at the actual intensity and detuning  \cite{Metcalf1999book} and the branching ratio $b$ for decay from the  excited state to all the possible metastable states \cite{Mcclelland2006lcw}, we estimate a decay rate $1/\tau=b \Gamma_a$ of the order of  $\unit[0.1]{s^{-1}}$.

\begin{figure}[t]
\includegraphics[width=1\columnwidth] {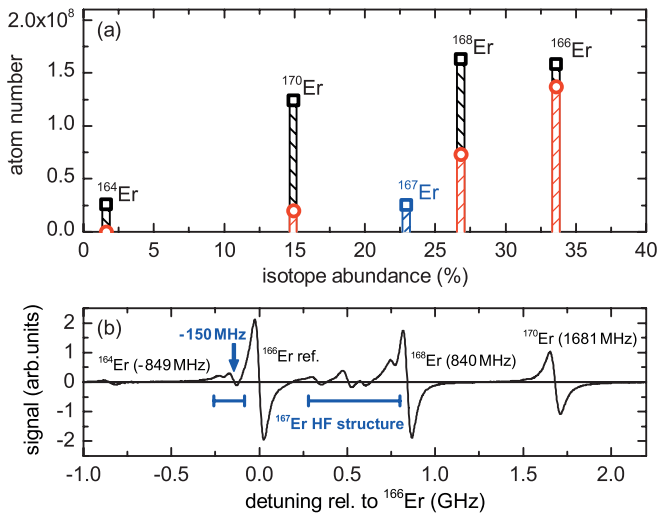}
\caption{(color online). Behavior of different Er isotopes. (a) MOT atom number for the four most abundant bosonic isotopes as well as for the $^{167}$Er fermionic isotope is plotted as a function of the natural abundance for $\unit[20]{s}$ (squares) and  $\unit[5]{s}$ (circles) loading time. (b) Spectroscopy signal of the blue transition shows the isotope shifts for the bosonic isotopes and the HF structure of $^{167}\mathrm{Er}$. The arrow indicates the locking point used for the ZS light to produce the $^{167}\mathrm{Er}$ MOT.}
\label{fig:isotopes}
\end{figure}

We could also demonstrate trapping of all the other Er isotopes, with the exception of the rare $^{162}\mathrm{Er}$ ($0.1\%$ natural abundance).
For all the bosonic isotopes we used about the same values for the detuning of the ZS and the MOT light. Figure \ref{fig:isotopes}(a) shows the atom number in the MOT for the different isotopes as a function of their natural abundance.
For a long loading time of $\unit[20]{s}$ we observe similar atom numbers exceeding $10^8$ for the three most abundant bosonic isotopes, indicating that saturation effects might apply. For a short loading time of $\unit[5]{s}$, the atom number increases with the natural abundance. However, we  observe a more complicated behavior than the expected linear growth.
This might be due to slight differences in the optimal MOT parameters, but may also point to differences in scattering and collisional properties among the different isotopes.

For the fermionic isotope $^{167}\mathrm{Er}$ we observe a MOT with atom number of about $3\times 10^7$, which is substantially lower than the numbers measured for the bosonic isotopes. A simple explanation of this behavior can be that we decelerate and cool only atoms in the $F=19/2$ hyperfine state, which has a statistical weight of about $20\%$. Note that a similar behavior has been observed with the fermionic $^{161}\mathrm{Dy}$ MOT, which also shows lower atom numbers than the ones of bosonic MOTs \cite{Youn2010dmo}. An additional complication stems from the unknown hyperfine splitting of the 401-nm line. From the spectroscopic signal we could  not identify \emph{a priori} where to lock the Zeeman slower light to be resonant with the desired $F=19/2\rightarrow F'=21/2$ hyperfine transition. To produce the $^{167}\mathrm{Er}$  MOT, we had to proceed blindly by first locking the 583-nm MOT light on the cooling transition and by  then  trying different locking points for  the ZS light  until the MOT was visible. We finally succeed in creating a MOT by locking the ZS light to the spectroscopic line located 150~MHz below the blue transition frequency of the $^{166}\mathrm{Er}$ isotope; see the arrow in Fig.\,\ref{fig:isotopes}(b).

After loading the MOT we apply a stage of compression to reduce the temperature of the atomic cloud and to achieve good starting conditions for direct loading of an optical dipole trap. With the optimized parameters $\delta_{583}=\unit[-0.5]{\Gamma_{583}}$, $I=\unit[0.16]{I_{s, 583}}$, and $B'=\unit[0.8]{G/cm}$, we measure a temperature of  $\unit[15]{\mu K}$ via time-of-flight experiments.  In addition, Stern-Gerlach experiments indicate that atoms in the compressed MOT phase are naturally pumped in the lowest Zeeman sublevel $m_J=-6$ by the MOT light. This can be explained by considering a combined effect  of the narrow cooling transition used for the MOT and of gravity. The latter pushes the atoms downward creating an effective imbalance in the trap and leading to a preferential absorbtion of the $\sigma^-$ polarized light from the lower vertical beam. For $2\times 10^8$ atoms at $T=\unit[15]{\mu K}$ we estimate a peak number density of the polarized sample of $\unit[1.5 \times 10^{11}]{cm^{-3}}$, corresponding to a phase-space density of about $4\times 10^{-6}$.
These values are similar to the ones observed in Yb experiments using the intercombination light \cite{Kuwamoto1999mot}. Compared to other experiments on magnetic lanthanides, such as Dy \cite{Lu2011sdb}, our much simpler approach based on a single cooling light for the MOT provides higher atom numbers and similar final temperatures. We suggest that this scheme can be successfully implemented with Dy, Ho and Tm, using the 626-nm ($\Gamma/2\pi=135$~kHz) \cite{Lu2011soa}, the 598-nm ($\Gamma/2\pi=146$~kHz) \cite{Saffman2008stn}, and the 531-nm ($\Gamma/2\pi=370$~kHz) \cite{Sukachev2011lco} transitions, respectively.

In conclusion, we have demonstrated an efficient and simple approach for an Er MOT based on a single narrow-line transition. Our scheme works with all abundant Er isotopes and allows for direct loading of an optical dipole trap. In first experiments we were able to load up to $10^7$ Er atoms into the dipole trap in a single focused-beam configuration. Optimization of the dipole trap loading and evaporative cooling experiments are under way in our laboratory.

We are grateful to  J.\,J.\,McClelland and A.\,J.\,Berglund for sharing with us their precious knowledge on Er and to S.\,Kotochigova, O.\,Dulieu, M.\,Lepers, and J.\,F.\,Wyart for fruitful discussions. We also thank the Sr team in Innsbruck and the Yb team in Tokio for their support. This work is supported by the Austrian Ministry of Science and Research (BMWF) and the Austrian Science Fund (FWF) through a START grant under Project No.\,Y479-N20 and by the European Research Council under Project No.\,259435.

\bibliographystyle{apsrev}



\end{document}